\documentstyle[aps,prl,preprint,floats,epsfig]{revtex}

\begin{document}

\preprint{\tighten\vbox{\hbox{\hfil CLNS 01/1718}
                        \hbox{\hfil CLEO 01-02}
}}

\title{Search for $B^0\rightarrow\pi^0\pi^0$ decay}  

\author{CLEO Collaboration}
\date{March 23, 2001}

\maketitle
\tighten

\begin{abstract} 
We have searched for the charmless hadronic decay of $B^0$~mesons into
two neutral pions. Using $9.13\,{\rm fb}^{-1}$ taken at the
$\Upsilon(4{\rm S})$ with the CLEO detector, we obtain an improved
upper limit for the branching fraction ${\cal B}(B^0\rightarrow\pi^0\pi^0) <
5.7\times 10^{-6}$ at the 90\% confidence level.
\end{abstract}
\newpage

{
\renewcommand{\thefootnote}{\fnsymbol{footnote}}

\begin{center}
D.~M.~Asner,$^{1}$ A.~Eppich,$^{1}$ T.~S.~Hill,$^{1}$
R.~J.~Morrison,$^{1}$
R.~A.~Briere,$^{2}$ G.~P.~Chen,$^{2}$ T.~Ferguson,$^{2}$
H.~Vogel,$^{2}$
A.~Gritsan,$^{3}$
J.~P.~Alexander,$^{4}$
% ...
\end{center}
 
\small
\begin{center}
$^{1}${University of California, Santa Barbara, California 93106}\\
$^{2}${Carnegie Mellon University, Pittsburgh, Pennsylvania 15213}\\
$^{3}${University of Colorado, Boulder, Colorado 80309-0390}\\
$^{4}${Cornell University, Ithaca, New York 14853}\\
% ...
\end{center}

\setcounter{footnote}{0}
}
\newpage

$CP$~violation in the neutral kaon system was observed long
ago~\cite{pdg}, and evidence of $CP$~violation in $B$~decays is
beginning to be seen~\cite{cdf}.  In the standard model,
$CP$~violation arises naturally from a single complex phase in the
Cabibbo-Kobayashi-Maskawa~(CKM) quark-mixing matrix~\cite{ckm}.
Observation of the time-dependent $CP$-violating asymmetry in the
decay $B^0\rightarrow\pi^+\pi^-$ (charge-conjugate modes are implied
throughout this Letter) would, in principle, give a measurement of the
sum of the CKM phases $\beta\equiv\arg(V^*_{td})$ and
$\gamma\equiv\arg(V^*_{ub})$. However, difficulties arise from the
fact that the tree and penguin contributions to the
$B^0\rightarrow\pi^+\pi^-$ decay enter with similar amplitude and
unknown relative phase. It is known already from the large ratio of
branching fractions ${\cal B}(B\rightarrow K^+\pi^-)/{\cal
B}(B\rightarrow\pi^+\pi^-)$ that the penguin contribution is
large~\cite{pipi}. The tree and penguin contributions can (in
principle) be separated by performing an isospin analysis on the
related $B\rightarrow\pi\pi$ modes~\cite{gronau90}.  Although the
decay mode $B^0\rightarrow\pi^+\pi^-$ has been observed and there is
some indication for the mode
$B^+\rightarrow\pi^+\pi^0$~\cite{otherpipi}, the
$B^0\rightarrow\pi^0\pi^0$ decay mode has not been seen yet.
Theoretical calculations offer possible values for the branching
fraction ${\cal B}(B^0\rightarrow\pi^0\pi^0)$  ranging from $10^{-8}$
to $10^{-5}$~\cite{predictions}.

In this Letter, we present a new limit on the
$B^0\rightarrow\pi^0\pi^0$ branching fraction based on data taken with
the CLEO~II detector at the Cornell Electron Storage Ring (CESR). The
data consist of $9.13\,{\rm fb}^{-1}$ taken at the
$\Upsilon(4{\rm S})$, corresponding to $9.67\times 10^6 B\bar{B}$
pairs,\footnote{We assume equal branching fractions for
$\Upsilon(4S)\rightarrow B^0\bar{B}^0$ and $B^+B^-$.} and
$4.35\,{\rm fb}^{-1}$ taken below the $B\bar{B}$ threshold, used
for background studies. The new result supersedes the result from a
previous publication~\cite{CLEO_godang98}, which was obtained with one
third of the present statistics.

CLEO~II\ is a general purpose detector, described in detail
elsewhere~\cite{CLEO_det}. Most relevant for the present analysis are
the tracking system and the electromagnetic calorimeter.  Momentum and
specific ionization ($dE/dx$) of charged tracks are measured in
cylindrical drift chambers in a $1.5\,$T solenoidal magnetic field. In
a second configuration of the detector, CLEO~II.V, the innermost
tracking chamber was replaced by a 3-layer, double-sided silicon
vertex detector, and the gas in the main drift chamber was changed
from argon-ethane to a helium-propane mixture.  These modifications
improved both the charged particle momentum resolution and the $dE/dx$
resolution.  Photons are detected using a high resolution crystal
CsI(Tl) electromagnetic calorimeter, composed of $7800$-CsI(Tl)
crystals covering $96\%$ of the solid angle.
Approximately two thirds of the data used in the present
analysis was taken with the CLEO~II.V\ detector.

Photon candidates are defined as isolated showers with energy greater
than $30\,$MeV in the central region of the calorimeter ($|\cos \theta
|<0.71\,$, where $\theta$ is the polar angle relative to the beam
axis), and with energy greater than $50\,$MeV elsewhere. Neutral pions
are formed from pairs of isolated photons with invariant mass within
$2.5\,\sigma$ of the $\pi^0$ mass ($\sigma\approx 9\,{\rm MeV}/c^2$
for a $2.7\,{\rm GeV}/c$ $\pi^0$).  We require at least one of the two
photons forming a $\pi^0$ candidate to be in the central region of the
calorimeter.  The energies of the selected photons are then
kinematically fitted with the mass constrained to the $\pi^0$ mass.

The $B$ decay candidates are selected using a beam-constrained
$B$~mass $M=\sqrt{E^2_b-p^2_B}$, where $E_{\rm beam}$ is the beam
energy and $p_B$ is the $B$~candidate momentum, and an energy
difference $\Delta E=E_1+E_2-E_{\rm beam}$, where $E_1$ and $E_2$ are
the energies of the two neutral pions. The resolution in $M$ is about
$3.4\,{\rm MeV}/c^2$, due to equal contributions from the beam energy
spread and the $\pi^0$ energy resolution. The resolution on $\Delta E$
is approximately $60\,{\rm MeV}$ and is slightly asymmetric because of
energy loss out of the back of the CsI\ crystals. Using events
containing at least three charged tracks, we select $B$~candidates
with $M$ in the range $5.2-5.3\,{\rm GeV}/c^2$ and $|\Delta E| <
400\,{\rm MeV}$.  The fiducial region in $M$ and $\Delta E$ includes
the signal region and a substantial sideband used for background
normalization.

The main background arises from $e^+e^-\rightarrow q\bar{q}$ (where
$q=u,d,s,c$). Such events typically exhibit a two-jet structure and
can produce high momentum back-to-back particles (tracks and/or
showers) in the fiducial region. To reduce contamination from these
events, we calculate the angle $\theta_S$ between the sphericity
axis~\cite{wu84} of the candidate showers and the sphericity axis of
the rest of the event. The distribution of $\cos\theta_S$ is strongly
peaked at $\pm 1$ for $q\bar{q}$~events, due to their two-jet
structure, and is nearly flat for $B\bar{B}$~events. We require
$|\cos\theta_S|<0.8$ which eliminates $85\%$ of the
$q\bar{q}$~background. Using a detailed GEANT-based Monte~Carlo
simulation\cite{geant} we determined the overall
$B^0\rightarrow\pi^0\pi^0$ signal detection efficiency ${\cal E} =
28.8\,\%$, dominated by geometric acceptance. Additional
discrimination between signal and $q\bar{q}$~background (continuum) is
obtained from event shape information used in a Fisher discriminant
(${\cal F}$) technique as described in detail in
Ref.\cite{CLEO_asner96}.  The Fisher discriminant is used in the
maximum-likelihood fit described below.

\begin{figure}[hbt]
\centering
\leavevmode
\epsfxsize=3.25in
\epsffile{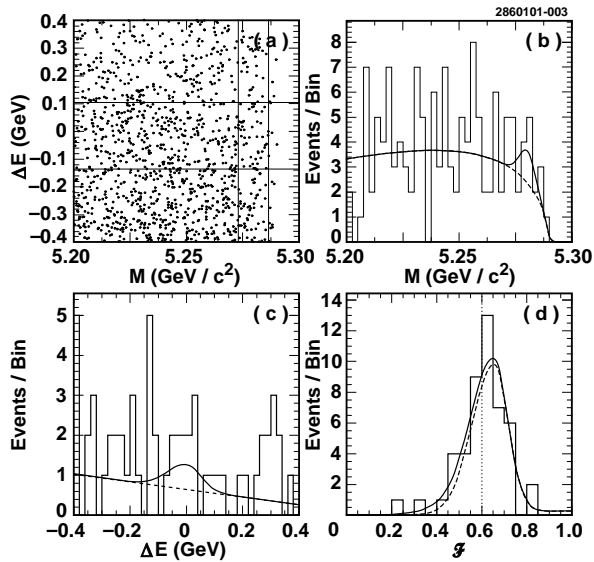}
\caption{Distributions of the selected events.  a)~$\Delta E$ versus
$M$ for all selected events (only $|\cos\theta_S|<0.8$ restriction is
applied). The solid lines show the $2\,\sigma$ boundaries.  b)~$M$
distribution after~$2\sigma$ requirements on $\Delta E$ and ${\cal F}
< 0.6$.  c)~$\Delta E$ distribution after~$2\sigma$ requirements on
$M$ and ${\cal F} < 0.6$.  d)~${\cal F}$ after~$2\,\sigma$
requirements on $\Delta E$ and $M$.  The dotted line shows the
position of the cut on~${\cal F}$ for the other two variables. In
plots~b), c) and~d), the solid line shows the result of the fit for
signal and background, and the dashed line shows the contribution of
the background alone.}
\label{fig:one}
\end{figure}

A total of $1134$ $B^0\rightarrow\pi^0\pi^0$ candidates are selected
with the requirements described above. Figure \ref{fig:one} shows the
$\Delta E$ versus $M$ distribution of all the selected events, and the
individual distributions of $M$, $\Delta E$ and ${\cal F}$, with
restrictions on the other two variables to emphasize the signal region.


Monte~Carlo simulation was also used to study backgrounds from
$b\rightarrow c$ and other $b\rightarrow u$ and $b\rightarrow s$
decays. More than $40$~decay modes of the $B$~meson into final states
containing energetic $\pi^0$s and/or photons were considered.  Only
the $B^+\rightarrow\rho^+\pi^0$ decay channel was found to give a
non-negligible contribution to the selection of
$B^0\rightarrow\pi^0\pi^0$ signal events.  The three-body final state
$\pi^+\pi^0\pi^0$ can be misidentified as a two-body $\pi^0\pi^0$
signal candidate when the charged pion from the asymmetric decay of
the polarized~$\rho^+$ has sufficiently low momentum and the
$\pi^0$~energies are poorly measured.  The best separation between
this background and signal is obtained in the $\Delta E$ distribution.
The $B^+\rightarrow\rho^+\pi^0$ background is accounted for in the
maximum-likelihood fit as described below.

To extract the signal yield, we perform an unbinned maximum-likelihood
fit using the variables $M$, $\Delta E$, and ${\cal F}$ for each
candidate event. The likelihood of an event is parameterized by the
sum of probabilities of signal, $q\bar{q}$~background, and
$B^+\rightarrow\rho^+\pi^0$ background hypotheses, with relative
weights determined by maximizing the likelihood function~${\cal L}$.
The probability of a particular hypothesis is calculated as the
product of the probability density functions (PDFs) for each of the
input variables.  Further details about the likelihood fit can be
found in Ref.~\cite{CLEO_asner96}. The PDFs for signal and
$B^+\rightarrow\rho^+\pi^0$ are determined from high-statistics
Monte~Carlo samples. The PDFs for continuum are obtained from the data
taken below the $B\bar{B}$ threshold.

Monte~Carlo experiments are generated to test the fitting procedure,
and to produce frequentist confidence intervals as defined in
Ref.~\cite{pdg}.  We generate Monte~Carlo samples containing the same
number of events as the real data sample. Continuum events are
generated according to the continuum PDFs, neglecting the small
correlation between the fit variables. According to a given branching
fraction ${\cal B}(B^0\rightarrow\pi^0\pi^0)$, we include signal
events randomly selected from our large $B^0\rightarrow\pi^0\pi^0$
Monte~Carlo simulated sample. We also include Monte~Carlo simulated
$B^+\rightarrow\rho^+\pi^0$ events. We generate $1000$~samples for
several values of ${\cal B}(B^0\rightarrow\pi^0\pi^0)$ and ${\cal
B}(B^+\rightarrow\rho^+\pi^0)$ in the range $0-10\times 10^{-6}$ and
$0-42\times 10^{-6}$, respectively.  We apply the fitting procedure to
every sample individually and determine the signal yield distribution
for each value of ${\cal B}(B^0\rightarrow\pi^0\pi^0)$ and ${\cal
B}(B^+\rightarrow\rho^+\pi^0)$.  In the samples containing
$B^+\rightarrow\rho^+\pi^0$ events, we find a small increase of the
signal yield proportional to ${\cal B}(B^+\rightarrow\rho^+\pi^0)$. At
the $90\%$ confidence level (C.L.) upper limit ${\cal
B}(B^+\rightarrow\rho^+\pi^0) < 42\times
10^{-6}$~\cite{CLEO_jessop00}, the contribution to the signal yield is
$0.3$ event.  We include this maximal contribution as a one-sided
systematic uncertainty in the result.  The Monte~Carlo experiments
show that once the $B^+\rightarrow\rho^+\pi^0$ background is accounted
for, the average yield for any value of ${\cal
B}(B^0\rightarrow\pi^0\pi^0)$ is equal to the expected yield for this
branching fraction, excluding thus any significant bias from the
fitting method. An estimation of the statistical sensitivity of the
measurement is given by the width of the yield distributions, and is
measured to be about $\pm 5$~events.

\begin{figure} [hbt]
\begin{center}
\epsfxsize=3.25in
\epsffile{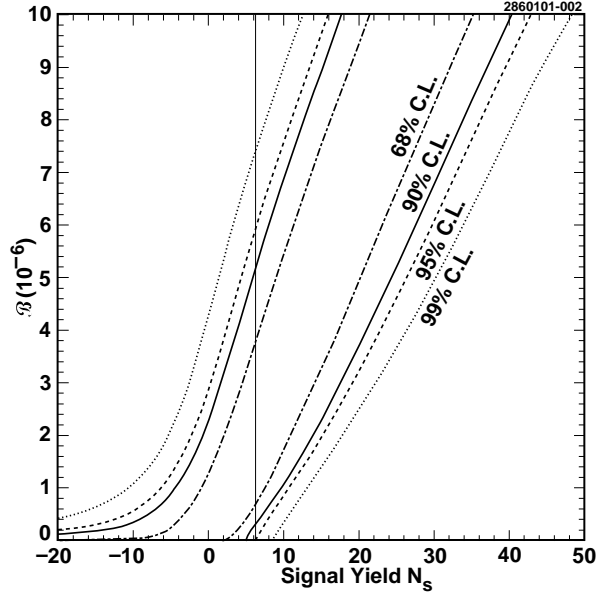}
\caption{Frequentist confidence intervals for the branching fraction
${\cal B}(B^0\rightarrow\pi^0\pi^0)$ versus signal yield~$N_S$, as
determined from Monte~Carlo experiments. No systematic effects are
included. The vertical line indicates the likelihood fit result for
$N_S$.}
\label{fig:interval_belt}
\end{center}
\end{figure}

Figure~\ref{fig:interval_belt} shows the $68\%$, $90\%$, $95\%$,
and $99\%$ frequentist confidence intervals (statistical only) built
from the signal yield distributions obtained with ${\cal
B}(B^+\rightarrow\rho^+\pi^0)=0$ and following the method described in
Ref.~\cite{feldman98}.

\begin{figure} [hbt]
\begin{center}
\epsfxsize=3.25in
\epsffile{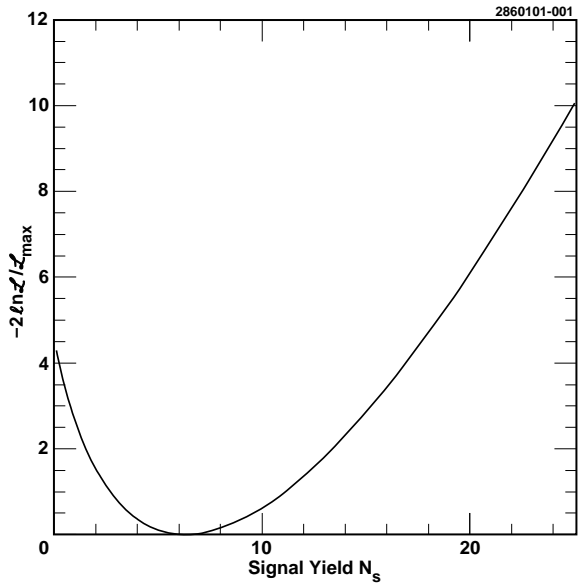}
\caption{Likelihood function $-2\ln{\cal L}/{\cal L}_{\rm max}$ versus
$B^0\rightarrow\pi^0\pi^0$ signal yield.}
\label{fig:chi2}
\end{center}
\end{figure}

Figure~\ref{fig:chi2} shows the result of the likelihood fit as a plot
of $\chi^2-\chi^2_{\rm min} = -2\ln{\cal L}/{\cal L}_{\rm max}$. The
maximum likelihood ${\cal L}_{\rm max}$ is found for a signal yield
$N_S=6.2^{+4.8}_{-3.7}$ events, with a statistical significance of
$2.0\,\sigma$.  We define the  statistical significance to be
$n\sigma$ if the value of $-2\ln{\cal L}$ increases by $n^2$ when the
signal yield $N_S$ is  constrained to be zero. The measured yield for
$B^+\rightarrow\rho^+\pi^0$ is $N_{\rho\pi}=-11\pm 9$ events,
consistent with the upper limit for that
mode~\cite{CLEO_jessop00}. This yield should, however, not be used to
calculate a new value of the upper limit for ${\cal
B}(B^+\to\rho^+\pi^0)$, as the $\pi^0\pi^0$ analysis described here is
deliberately designed to minimize  sensitivity to the
$B^+\to\rho^+\pi^0$ mode. We also checked that, due to the small
correlation between signal and background, the signal yield is reduced
by only $0.2$ event when the background yield is constrained to be
positive.  Given the measured value of $N_S$, we use
Figure~\ref{fig:interval_belt} to determine the frequentist confidence
interval for  ${\cal B}(B^0\rightarrow\pi^0\pi^0)$. We obtain the
$90\%$ C.L.\ interval $0.3\times 10^{-6}<{\cal
B}(B^0\rightarrow\pi^0\pi^0)<5.2\times 10^{-6}$.  This interval is
statistical only, and does not include systematic uncertainties.

For the treatment of systematic uncertainties, we separate them into
two categories. First, we estimate a systematic uncertainty on the
fitted signal yield by varying the PDFs used in the fit within their
uncertainties, and we add in quadrature the previously described
systematic uncertainty due to the possible residual contamination from
$B^+\rightarrow\rho^+\pi^0$ events.  We also consider a possible
mismodeling of the $e^+e^-\rightarrow\tau^+\tau^-$ contribution in the
PDFs for continuum, and we assign a systematic uncertainty of $\pm 0.7$
event based on a high-statistics Monte~Carlo simulation study.  We
obtain the total systematic uncertainty ${}^{+2.0}_{-1.8}$~events on
the signal yield~$N_S$. Secondly, we estimate an uncertainty on the
signal detection efficiency ${\cal E}$, to account for uncertainties
related to $\pi^0$ finding efficiency, maximum-likelihood fit
efficiency, luminosity, Monte~Carlo statistics, and the
$|\cos\theta_S|$ requirement. The efficiency with its uncertainty is
${\cal E} = (28.8\pm 2.0)\%$.

We derive the central value of the branching fraction ${\cal B}
(B^0\rightarrow\pi^0\pi^0) = \left(2.2
^{+1.7}_{-1.3}\,{}^{+0.7}_{-0.7} \right)\times 10^{-6}$, where the
first uncertainty is statistical and the second is systematic.  We
also calculate the $90\%$ C.L.\ upper limit yield by integrating the
likelihood function
\begin{displaymath}
\frac{\int_0^{N^{UL}}{\cal L}_{\rm max}(N)\,dN}
     {\int_0^{\infty}{\cal L}_{\rm max}(N)\,dN} = 0.90,
\end{displaymath}
where ${\cal L}_{\rm max}$ is the maximum ${\cal L}$ at fixed $N$ to
conservatively account for possible correlations among the free
parameters in the fit. We obtain the upper limit $N_S<13.7$ events at
$90\%$ C.L. (statistical). We then calculate the corresponding upper
limit of the branching fraction, add one standard deviation of the
systematic uncertainty, and obtain the branching fraction $90\%$
C.L. upper limit ${\cal B}(B^0\rightarrow\pi^0\pi^0) < 5.7\times
10^{-6}$.

In summary, using the full CLEO~II\ and CLEO~II.V\ data set, we have
obtained an improved upper limit on the branching fraction of the
$B^0\rightarrow\pi^0\pi^0$ decay mode. We see no indication for a
signal and we set a new $90\%$ C.L. upper limit ${\cal
B}(B^0\rightarrow\pi^0\pi^0) < 5.7\times 10^{-6}$.  This limit is in
the range of the theoretical predictions and constrains some of them.

We gratefully acknowledge the effort of the CESR staff in providing us with
excellent luminosity and running conditions.
This work was supported by 
the National Science Foundation,
the U.S. Department of Energy,
the Research Corporation,
the Natural Sciences and Engineering Research Council of Canada, 
the Swiss National Science Foundation, 
the Texas Advanced Research Program,
and the Alexander von Humboldt Stiftung.  



\end{document}